# Controllable spin splitting in 2D Ferroelectric few-layer γ-GeSe

Shuyi Shi, [1] Kuan-Rong Hao, [2] Xing-Yu Ma, [2] Qing-Bo Yan [1]* and Gang Su [3, 1, 2]*

1. Center of Materials Science and Optoelectronics Engineering, College of Materials Science and Optoelectronic Technology, University of Chinese Academy of Sciences, Beijing 100049, China.

2. School of Physical Sciences, University of Chinese Academy of Sciences, Beijing 100049, China.

3. Kavli Institute for Theoretical Sciences, and CAS Center of Excellence in Topological Quantum Computation, University of Chinese Academy of Sciences, Beijing 100190, China.

* Email: yan@ucas.ac.cn; gsu@ucas.ac.cn

# Abstract

γ-GeSe is a new type of layered bulk material that was recently successfully synthesized. By means of density functional theory first-principles calculations, we systematically studied the physical properties of two-dimensional (2D) few-layer γ-GeSe. It is found that few-layer γ-GeSe are semiconductors with band gaps decreasing with increasing layer number; and 2D γ-GeSe with layer number n≥2 are ferroelectric with rather low transition barriers, consistent with the sliding ferroelectric mechanism. Particularly, spin-orbit coupling induced spin splitting is observed at the top of valence band, which can be switched by the ferroelectric reversal; furthermore, their negative piezoelectricity also enables the regulation of spin splitting by strain. Finally, excellent optical absorption was also revealed. These intriguing properties make 2D few-layer γ-GeSe promising in spintronic and optoelectric applications.

# Introduction

Group IV-VI monochalcogenides (GeS, GeSe, SnS, SnSe, etc.) can form various polymorphs with different geometric structures and exhibit intriguing physical, electric, and chemical properties. Their most common α-phase, which has an atomically puckered layered structure like black phosphorus, along with their corresponding 2D counterparts, has attracted wide attention and extensive studies in recent years [1-6]. Spontaneous in-plane ferroelectricity [7] and ferroelasticity had been found in monolayer group IV monochalcogenides, making them novel 2D multiferroics [8,9]. Diverse anisotropy of their phonon transport has also been revealed [10]. Not only bulk SnSe had been found to possess excellent thermoelectric properties [11,12], but also monolayer SnSe had been predicted to have ultralow lattice thermal conductivity and high hole mobility, and thus could have potential applications in thermoelectrics and optoelectronics [13]. In 2017, another polymorph, i.e., the β-phase of bulk GeSe with a boat conformation has also been successfully synthesized [14], where the monolayer β-GeSe has been predicted to have anisotropic electric transport properties, high electron mobility [15], spontaneous in-plane ferroelectricity [16], and strong optical absorbance induced by saddle-point exciton absorptions [17].

Recently, a new polymorph of GeSe (γ-GeSe) was successfully synthesized and characterized [18], which has a hexagonal layered structure and exhibits high electrical conductivity. It agrees well with the predictions from first-principles by Zou et al. [19], which demonstrated the large low energy absorbance induced by the camel back-like band edge structure, and suggested its potential in achieving high-temperature electron–hole liquid. The synthesis of bulk γ-GeSe stimulated research interests in its 2D counterpart [20]. The effects of strain and electric field on electronic and optical properties of monolayer γ-GeX (X= S, Se, and Te) [21,22] and the interface contact with metallic systems [23] have been studied. It has also been predicted that monolayer γ-GeSe has low thermal conductivity [24] and is promising in thermoelectric applications [25,26]. Layer stacking can strongly affect interlayer interaction and symmetry of 2D materials, and novel physical properties may emerge. The thickness-dependent vibrational and electronic properties induced by strong interlayer interactions have been revealed in few-layer black phosphorus [27,28]. Interlayer excitons in 2D hetero-bilayers [29], especially in transition metal dichalcogenides (TMDCs), have aroused extensive studies [30,31]. Bilayer WSe$_2$ was proposed to be a natural platform for interlayer exciton condensates [32].

Unconventional superconductivity and other emergent properties have been discovered in magic-angle twisted graphene bilayers [33]. Spontaneous out-of-plane electric polarization that can be switched by translational sliding has been found in few-layer WTe$_2$ [34,35], 2D binary compounds such as bilayer BN [36,37,38], rhombohedral-stacked bilayer TMDCs [39], twisted bilayer TMDCs [40], and untwisted heterobilayer TMDCs [41], etc., which originate from interlayer vertical charge transfer and are referred as interfacial or sliding ferroelectricity. Coupled ferroelectricity and superconductivity in bilayer T$_d$-MoTe$_2$ were also reported very recently [42].

In this paper, we systematically investigated the few-layer γ-GeSe by first-principles calculations and focused on the novel properties arising from layer stacking. It is found that few-layer γ-GeSe are semiconductors with band gaps decreasing with increasing the layer number. The 2D γ-GeSe with layer number n≥2 breaks the inversion symmetry that the monolayer possesses and gives rise to emerging ferroelectricity and piezoelectricity. The ferroelectric transition path of bilayer γ-GeSe has been revealed with rather low reversal barriers and can be explained by the sliding ferroelectric mechanism. Interestingly, spin-orbit coupling induced spin splitting is observed at the valence band top, which can be switched by the ferroelectric reversal and tuned by strain due to their negative piezoelectricity, and thus is controllable. We also unveiled high carrier mobility and excellent optical absorption of bilayer γ-GeSe. All these findings make the few-layer γ-GeSe a promising candidate for applications in spintronic and optoelectric devices.

## Methods

The first-principles density functional theory (DFT) calculations are performed with the projector augmented wave pseudopotentials as implemented in the Vienna *ab initio* simulation package (VASP) [43][44]. The Perdew-Burke-Ernzerhof parametrization of generalized gradient approximation (GGA) is used for the exchange-correlation functional [45]. The plane-wave cutoff energy is 500 eV. The 9×9×1 k-point grid is adopted in both structural relaxation and static calculations. To describe the interlayer interaction, the van der Waals correction is implemented using the DFT-D2 method [46]. The hybrid functional HSE06 is used to accurately calculate the electronic structures [47]. A vacuum region of 30 Å thickness is set to avoid the interaction between the periodic images. The energy convergence criterion is set to $10^{-6}$ eV. The force

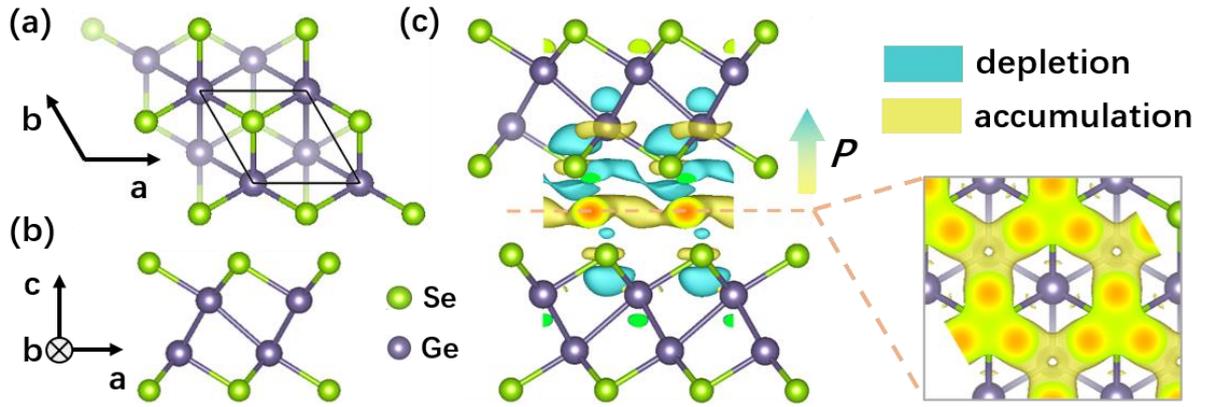

**Figure 1** (a) Top view and (b) side view of the geometric structure of monolayer γ-GeSe. The green and grey balls represent Se and Ge atoms, respectively. The parallelogram marks out the unitcell. (c) The geometric structure and charge density difference diagram of bilayer γ-GeSe. The cyan and yellow colours indicate the charge depletion and accumulated regions, respectively, and the isovalue is $10^{-4}$ e/bohr$^3$. The arrow with *P* indicates the direction of electric polarization. The cross-section of charge accumulation area in the gap between two layers are featured in the right panel.

converge criterion during the structural relaxation is 0.001 eV/Å. The out-of-plane electric polarization was calculated by integrating the charge density times the position vector, which has also been adopted in previous works [35][48][49]. The climbing image nudged elastic band (CI-NEB) method [50] is used to simulate the ferroelectric transition path. The spin texture maps are obtained by using VASPKIT [51]. The optical absorption spectra are obtained using the GW approximation and the Bethe-Salpeter equation (BSE) method [52][53]. More details are presented in supplementary materials.

## Results and discussion

### The geometric structure of 2D few-layer γ-GeSe

The conventional α-GeSe has a puckered layered structure similar to that of the isoelectronic black phosphorus. Although the Ge and Se atoms are sp$^3$-like bonded and form six-membered rings with "chair" conformation, the lattice is orthorhombic and each layer is two-atom thick due to the puckering. β-GeSe is also made up of buckling two-atom thick layers, but the six-membered rings are in the 'boat' conformation [14]. The geometric structure of γ-GeSe is also layered but is distinctly different from the two polymorphs. As shown in Fig. 1(a)(b), monolayer γ-GeSe has a hexagonal lattice with a space group of *P-3m1* (No. 164) and point group of $D_{3d}$, indicating it is centrosymmetric with a geometric inversion center, and thus has no electric polarization. It is four-atom-thick (Se-Ge-Ge-Se) and can be viewed as two merged buckled honeycomb sublayers with similar structure of blue phosphorus [54]. Fig. 1(c) shows the geometric structure of the bilayer γ-GeSe. We have compared AB and AB' stackings, and the results show that AB' stacking is more stable than AB stacking, in good agreement with previous work on bulk γ-GeSe [18]. Thus, here we only consider AB' stacking. Interestingly, while monolayer γ-GeSe is centrosymmetric, the AB' stacking bilayer γ-GeSe loses inversion symmetry, that is, the space group and point group change to *P3m1* (No. 156) and $C_{3v}$, respectively, implying that ferroelectricity may emerge, which will be discussed in detail below.

## The ferroelectric properties and transition path

Fig.1(c) shows the interlayer charge redistribution of the bilayer γ-GeSe, which is evaluated by the charge density difference between the whole bilayer and two separated monolayers. The cyan and yellow colours represent the charge-depleted regions and accumulated regions, respectively. Interestingly, although the geometric structures of the bottom layer and top layer are identical, the corresponding charge distributions are different, and distinct positive and negative charge separation can be observed. As featured in the right panel of Fig. 1(c), a remarkable high-density region of charge accumulation (yellow) in the gap between upper and lower GeSe layers composes a hexagonal negative charge area, which mainly locates at the top of Se atoms in the bottom layer. On the other hand, the high-density regions of charge depletion (blue) are located around the lower Se atoms in the upper layer. Thus, it clearly shows a charge transfer from the upper layer to the bottom layer, which gives rise to an out-of-plane electric polarization pointing from the bottom layer to the upper layer as indicated in Fig. 1(c).

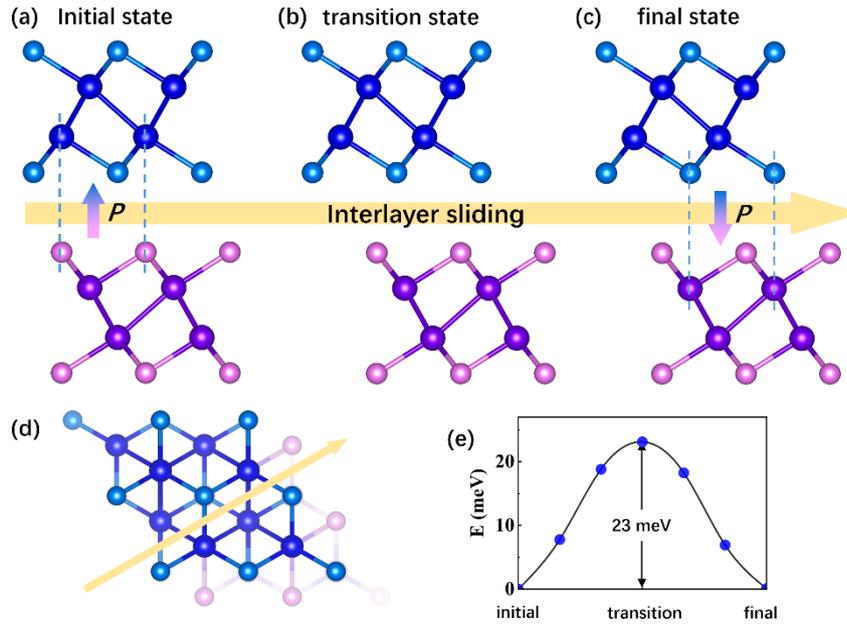

**Figure 2** The ferroelectric transition illustration of bilayer γ-GeSe. (a), (b) and (c) show the side views of the initial, transition and final states during the transition, respectively. The upper and lower layers are distinguished in blue and purple, respectively. The short arrows with *P* indicate the direction of electric polarization. The long yellow arrow indicates the interlayer sliding direction. The blue dashed lines indicate the relative positions of the upper and lower atoms. (d) Top view of the initial state in bilayer γ-GeSe. The yellow arrow also indicates the interlayer sliding direction. (e) shows the energy variation during the ferroelectric transition with a reversal barrier of 23 meV.

Now we calculate the out-of-plane electric polarization, and the obtained values are 0.047 μC/cm² (0.586 pC/m in 2D unit) and 0.063 μC/cm² (1.266 pC/m in 2D unit) for bilayer and trilayer γ-GeSe, respectively, which are in the same order of magnitude as other 2D ferroelectric bilayer chalcogenides, such as the bilayer $WTe_2$ (0.051μC/cm²) [35], bilayer $MoS_2$ (0.97 pC/m), bilayer InSe (0.24 pC/m) and bilayer GaSe (0.46 pC/m) [36]. Note that all the latter have vertical electric polarization that can be switched by in-plane interlayer sliding, implying that few-layer γ-GeSe may possess similar ferroelectric mechanism.

Using the CI-NEB method, the ferroelectric switching pathway of bilayer γ-GeSe was studied and illustrated in Fig. 2. Figs. 2(a) and (c) indicate the initial and final state with polarization up and down, respectively, while Fig. 2(b) represents the medium transition state without polarization. The upper and lower layers are marked by purple and blue, respectively. The polarization reverse process can be viewed as the bottom layer is sliding relative to the top layer along the direction simultaneously perpendicular to the ***b*** and ***c*** directions, as indicated by the horizontal light-yellow arrow. Fig. 2(d) shows the top view of Fig. 2(a), in which the sliding direction is also displayed. Thus, the ferroelectric transition of the bilayer γ-GeSe can be explained by the sliding ferroelectric mechanism. [36,55] The energy profile of the transition path is shown in Fig. 2(e), from which the minimum transition barrier can be read as 23 meV/unitcell. It is comparable to other 2D chalcogenides sliding ferroelectrics such as bilayer $MoS_2$ (15 meV/unitcell) [56,57], but much lower than 2D displacement ferroelectric monolayer $In_2Se_3$ (66 meV/unitcell) [58]. Since the ferroelectricity of few-layer $MoS_2$ has been observed in experiment, [57] the ferroelectricity in the bilayer γ-GeSe may also be detectable experimentally.

**Layer-dependent electronic properties**

Fig. 3 (a)(b)(c) shows the electronic energy bands of the few-layer γ-GeSe obtained by PBE calculations. The results in monolayer γ-GeSe are consistent with previous work [19,20]. As indicated in Fig. 3(a), the conduction band minimum (CBM) is at Γ, while the valence band forms a camel's back-like structure around Γ. The hump $K_1$ along Γ-K is slightly higher than the hump $M_1$ along Γ-M, thus the valence band maximum (VBM) is at $K_1$, indicating that monolayer γ-GeSe is an indirect semiconductor. Zou et al. have predicted that the band nesting effect near the camel's back-like structure can induce a large excitonic absorbance [19]. Fig. 3(b) and 3(c) show the energy bands of bilayer and trilayer γ-GeSe, respectively. Both the valence band and conduction band split into two-fold or three-fold degenerated bands, and the VBM shifts up while CBM moves down, leading to a general trend that the energy gap decreases with the increase of the number of layers, which is due to the interlayer interactions [27]. The energy gaps of monolayer, bilayer, and trilayer γ-GeSe calculated at PBE level are 0.60, 0.35, and 0.17 eV, respectively. Their energy gaps at the Heyd-Scuseria-Ernzerhof (HSE06) [47] level are also obtained, *i.e.*, 0.95, 0.62, and 0.27 eV, respectively. Note that in trilayer γ-GeSe, the valence band at Γ shifts considerably and is very close to the VBM at $K_1$; while the downward shift of the conduction band around M is larger than that at Γ, and the CBM moves from Γ to a *k*-point near M. It has been reported that AB'-stacked bulk γ-GeSe also has an indirect band gap, where the VBM is at Γ and CBM is at a *k*-point near M. [20]

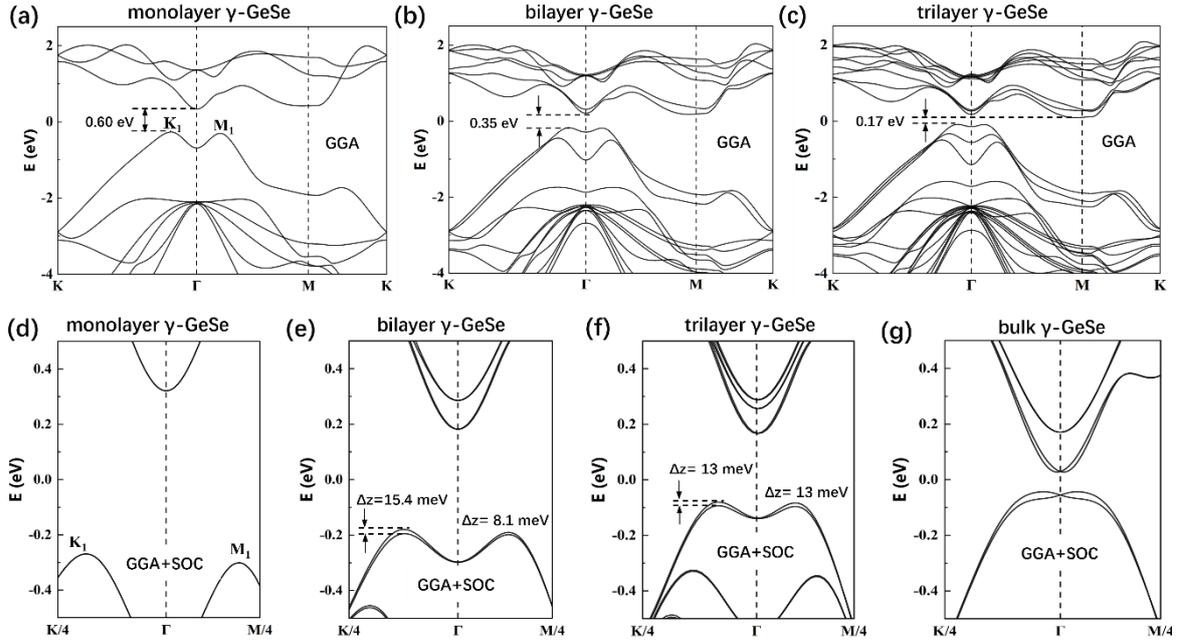

**Figure 3** Electronic structures of γ-GeSe few-layers. (a) (b) (c) Band structures for monolayer, bilayer and trilayer of γ-GeSe at GGA level, respectively. The corresponding band gaps are 0.6 eV, 0.35 eV and 0.17 eV, respectively. The valence band maxima are marked with $K_1$ and $M_1$; while $K_1$ is slightly higher than $M_1$. (d)(e)(f)(g) band structures for γ-GeSe monolayer, bilayer, trilayer and bulk γ-GeSe at GGA+SOC level, respectively. The spin splitting (Δz) at the top of valence band of bilayer γ-GeSe are indicated as 15.4 meV and 8.1 meV for $K_1$ and $M_1$, respectively.

## The switchable spin-orbit coupling induced spin splitting

Spin-orbit coupling (SOC) has not been considered in Fig. 3 (a)(b)(c). The energy bands including the SOC effect have also been calculated as shown in Fig. 3 (d)(e)(f). The energy band profiles and band gaps remain roughly the same. However, when focusing on the top valence bands of bilayer γ-GeSe as shown in Fig. 3(e), we find that the SOC lifts the spin degeneracy and the spin splitting occurs in the vicinity of $K_1$ and $M_1$. The splitting value at $K_1$ (15.4 meV) is much larger than that at $M_1$ (8.1 meV), exhibiting an asymmetric behavior along different paths in the Brillouin zone. No significant spin splitting is observed at Γ for both valence and conduction bands. As indicated in Fig. 3(f), the trilayer γ-GeSe also possesses similar spin splitting, and the splitting values are 13 meV at both $K_1$ and $M_1$, respectively. On the contrary, no spin splitting is observed in the monolayer γ-GeSe (Fig. 3(d)). Obviously, these splittings are strongly dependent on symmetry. As aforementioned, monolayer γ-GeSe is centrosymmetric and has an inversion center, while the bilayer and trilayer γ-GeSe break the inversion symmetry and possess spontaneous electric polarization. Similar spin splitting has also been reported in several nonmagnetic 2D and 3D systems [59,60,61], which may be a consequence of inversion symmetry breaking at non-time-reversal-invariant momentum (non-TRIM) [61]. We also calculate the GGA+SOC energy bands of bulk γ-GeSe and present them in Fig. 3(g), which is in good agreement with previous work [20]. Both the top valence band and bottom conduction band show Rashba-type spin splitting at Γ, which can be attributed to the noncentrosymmetric space group $P6_3mc$ (No. 186) and polar point group of $C_{6v}$ [18] of the bulk γ-GeSe, and Γ is a TRIM k-point.

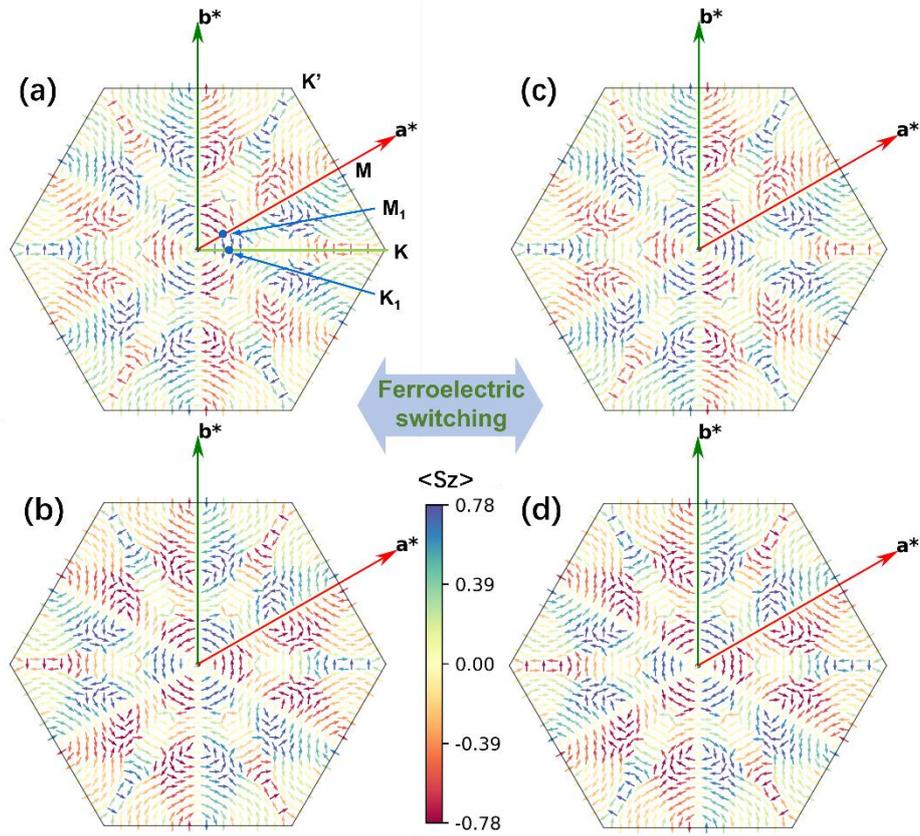

**Figure 4** The spin textures of valence bands of bilayer γ-GeSe at (a)(b) initial state (P↑) and (c)(d) final state (P↓) during ferroelectric switching, respectively. (a)(c) and (b)(d) correspond to the upper and lower spin-splitting valence bands, respectively. The red and green long arrows indicate the reciprocal lattice vectors. The short arrows represent the in-plane spin directions, and the colors indicate the out-of-plane spin. The $M_1$ and $K_1$ k-points are marked and located at about M/8 and K/8, respectively.

To delve deeper into the SOC effects, we calculate the momentum-dependent spin texture around Γ point of valence bands, as shown in Fig. 4(a)(b). The chiral in-plane spin texture can be observed around Γ point and the lower and upper valence bands have opposite spin directions and opposite in-plane chirality, clearly showing typical Rashba-like SOC characteristics. There are non-negligible out-of-plane spin components that cannot be explained by plain Rashba SOC, which could not explain the anisotropic $C_3$ symmetric feature of spin texture either, while the latter is consistent with the $C_{3v}$ geometric structure. Here we adopt an effective Hamiltonian with third-order terms (cubic k-terms) [62-64] as $H = v_k(k_x\sigma_y - k_y\sigma_x) + \lambda_k(k_x^3 - 3k_xk_y^2)\sigma_z$, where $v_k = v(1 + \alpha k^2)$ is the Dirac velocity with a second-order correction, α is the Rashba parameter, $\sigma_i$ are the Pauli matrices and $\lambda_k$ is the warping parameter. The *x* axis is along Γ-K. The first term is the Rashba term causing the in-plane chiral spin polarization, while the second term is invariant under $C_3$ rotation and is therefore responsible for the $C_3$ symmetric feature of spin texture. Besides, only the second term contains $\sigma_z$ and is hence responsible for the out-of-plane spin components. According to above Hamiltonian, the resulting spin vector $\mathbf{S} \propto [v_k\sin\theta, -v_k\cos\theta, \lambda_k\cos 3\theta]$, where $\theta = \tan(k_y/k_x)$ is the azimuth angle of mentum k with respect to the x-axis (Γ-K). Although the spin texture in the whole Brillouin zone is complicated, the above spin vector agrees well with both the in-plane and out-of-plane spin texture around Γ. Note that M₁ and K₁ k-points also locate in this area. Furthermore, the above Hamiltonian also leads to spin splitting $\Delta = \sqrt{v_k^2 k^2 + \lambda_k^2 k^6 \cos^2(3\theta)}$, where $k = \sqrt{k_x^2 + k_y^2}$. Thus, both Rashba term $v_k = v(1 + \alpha k^2)$ and third-order term contribute to the splitting.

Considering that the spin splitting has a strong correlation to the inversion symmetry breaking, which also leads to electric polarization in bilayer γ-GeSe and polarization direction can be reversed by ferroelectric switching, we speculate that the spin splitting may also be modulated by ferroelectric reversal. Then we obtained the spin texture around Γ point after ferroelectric switching, as shown in Fig. 4(c)(d). The arrows representing the in-plane spin polarization reverse directions before and after the ferroelectric transition, illustrating that the in-plane spin direction at VBM can be switched by the ferroelectric switching, which also suggests the possibility of tuning spin by electric field or gate voltage. In contrast, the color indicating the out-of-plane spin does not change before and after the ferroelectric transition. Recall that the first Rashba term and the second third-order term in Hamiltonian are responsible for the in-plane and out-of-plane spin components, respectively. The ferroelectric switching reverses the out-of-plane electric polarization and the sign of the Rashba term, but does not affect the third-order term, thus only in-plane spin is switched. Although the spin splitting values of the GeSe bilayer are somewhat small, there are various approaches to improve it, such as applying an external electric field, introducing heavy elements with strong atomic SOC, etc. We also find that strain can enhance spin splitting, as discussed below.

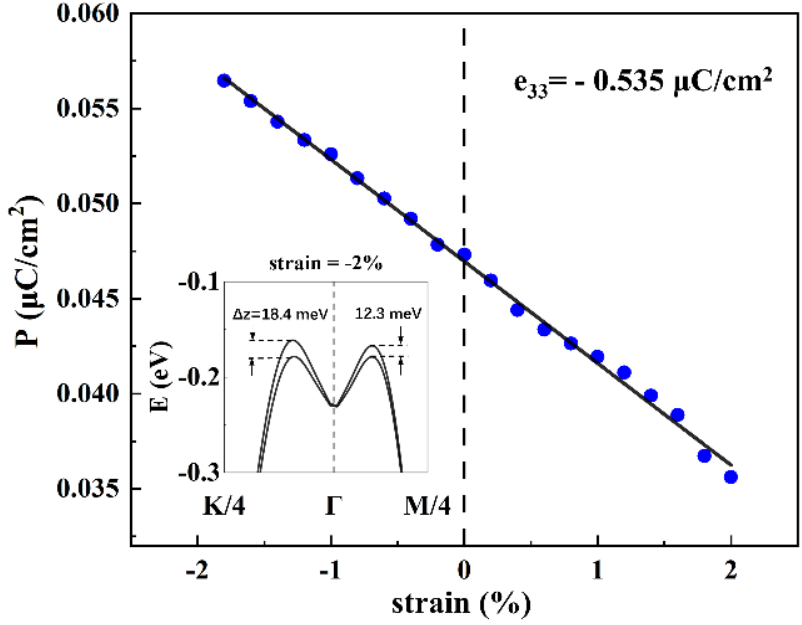

**Figure 5** The out-of-plane electric polarization as a function of strain along the c-axis for bilayer γ-GeSe, from which the piezoelectric coefficient e₃₃ can be derived. The positive and negative strains correspond to tensive and compressive strains, respectively. The inset shows the valence bands (GGA+SOC) of bilayer γ-GeSe at 2% compressive strain and the spin splitting values are indicated.

## The negative piezoelectricity and strain effect on spin splitting

Ferroelectric materials are usually piezoelectric materials. Since the bilayer γ-GeSe possesses out-of-plane ferroelectric polarization, here we mainly focus on the longitudinal piezoelectricity along ***c*** axis. The piezoelectric coefficient can be evaluated from the response of polarization to strain. As shown in Fig. 5, the polarization decreases with the increase of strain and the obtained value of piezoelectric coefficient $e_{33}$ is -0.535 μC/cm², which implies a negative piezoelectricity. Note that positive and negative strains correspond to tensive and compressive strains, respectively. A negative $e_{33}$ means that an out-of-plane compressive strain will enhance out-of-plane electric polarization, *i.e.*, pressure-enhanced ferroelectricity. The negative piezoelectric effect has also been discovered in other layered ferroelectrics [65,66,67,68]. The piezoelectric coefficient e$_{33}$ can be decomposed into a clamped-ion term calculated with internal atomic coordinates fixed at zero strain and internal-strain term arising from internal atomic relaxations in response to strain [65], which are -0.273 μC/cm² and -0.263 μC/cm², respectively. Thus, both the clamped-ion term and internal-strain term contribute almost half as much to the negative piezoelectricity of bilayer γ-GeSe.

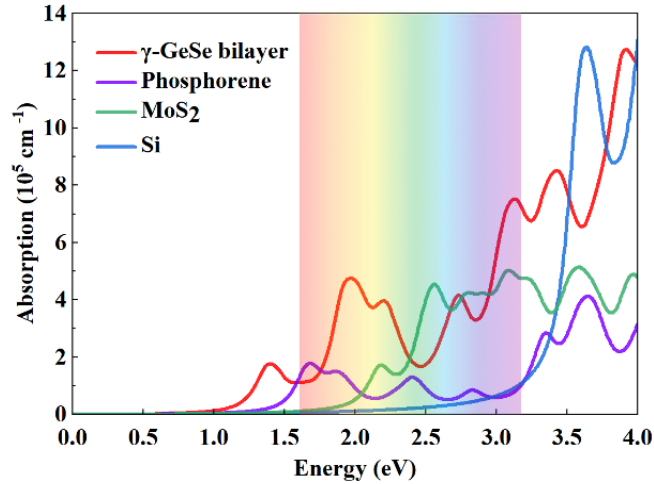

**Figure 6** The in-plane absorption spectra of γ-GeSe at GW level. Adsorption spectra of phosphorene, monolayer MoS$_2$ and Si are also included for comparison. The visible light range of solar spectra is also marked.

The negative piezoelectricity implies a potential way to modulate the spin splitting of bilayer γ-GeSe. The compressive strain will enhance the polarization and may also increase the spin-splitting values. To illustrate this conjecture, we calculated the energy bands at the PBE+SOC level at 2% compressive strain along the *c*-axis, as indicated in the inset of Figure 5. The obtained spin splitting at K$_1$ and M$_1$ are 18.4 meV and 12.3 meV, which means 19% and 52% enhancements under the compressive strain of 2%, respectively. Thus, the spin splitting of bilayer γ-GeSe cannot only be switched by ferroelectric reversal but also can be modulated by strain, exhibiting abundant regulation possibilities.

## The carrier's effective mass and optical absorption

The carrier effective mass of bilayer γ-GeSe can be calculated from the second derivative of CBM and VBM, which are 0.15m$_0$ and 0.42m$_0$ for electron and hole, respectively, where m$_0$ is the free electron mass. The small carrier's effective mass implies high carrier mobility. We also obtain the absorption spectra of bilayer γ-GeSe by using the GW+BSE method, as shown in Fig. 6, where the adsorption spectra of monolayer phosphorene, monolayer MoS$_2$, and Si are also included for comparison. The absorption profile covers the entire solar spectrum including the infrared and ultraviolet range, and the adsorption in the visible light range, especially from 1.5 to 2.5 eV, is distinctly higher than that of phosphorene and monolayer MoS$_2$, implying an outstanding solar radiation absorption. Therefore, bilayer γ-GeSe has a small effective carrier's mass and excellent solar radiation absorption, together with the out-of-plane ferroelectric polarization, which is helpful for the spatial separation of electrons and holes, suggesting its potential for electronic, photovoltaic, photodetector, and photocatalysis applications.

## Conclusions

In this paper, we systematically investigated the geometric and physical properties of few-layer γ-GeSe by first-principles calculations and pay close attention to the novel emerging properties induced stacking and interlayer interactions. The band gaps of few-layer γ-GeSe are found to decrease with increasing layer number; and the AB' stacking breaks the inversion symmetry of γ-GeSe with layer number n ≥ 2, giving rise to emerging ferroelectricity and piezoelectricity. The ferroelectric transition path of bilayer γ-GeSe has been revealed with rather low reversal barriers, and it turns out to be another new instance of sliding ferroelectrics. Interestingly, spin splitting is observed at the valence band top, which is a consequence of inversion symmetry breaking at non-TRIM k-points. Moreover, the above spin splitting can be switched by the ferroelectric reversal and tuned by strain as a result of their negative piezoelectricity, exhibiting various modulation possibilities. Besides, we also unveiled excellent optical absorption of bilayer γ-GeSe. All these findings make the few-layer γ-GeSe a promising candidate for applications in electronic, spintronic, and optoelectric devices.

# Acknowledgments

Q.B.Y. thanks Prof. Z.-G. Zhu, B. Gu, and Z.-C. Wang for helpful discussions. This work is supported in part by the National Key R&D Program of China (Grant No. 2018YFA0305800), the NSFC (Grant No. 12174386), the Chinese Academy of Sciences (CAS) Project for Young Scientists in Basic Research (YSBR-003), the CAS Information Plan (CAS-WX2021SF-0102), and the Fundamental Research Funds for the Central Universities. The calculations were performed at the Supercomputing Center of CAS and Beijing Super Cloud Computing Center.